%% file: d4d.tex
\documentclass[aps,prd,twocolumn,preprintnumbers,amsmath,amssymb,floats,nofootinbib]{revtex4} 
\usepackage{graphicx}
\usepackage{dcolumn}
\usepackage{bm}
\usepackage[english]{babel}
\usepackage[latin1]{inputenc} 
\usepackage{graphicx}
\usepackage{epstopdf}
\usepackage{amsfonts}
\usepackage{color}
\usepackage{epsfig}
\usepackage{mathrsfs}
\usepackage{verbatim} 
\usepackage{amsmath,amssymb}
\usepackage{subfigure}
\usepackage{hyperref}
\usepackage{framed}
\usepackage{lipsum}
\usepackage{tikz}
\usetikzlibrary{arrows,shapes}
\usetikzlibrary{arrows,automata}
\usetikzlibrary{mindmap,trees}

\makeatletter
\def\blfootnote{\xdef\@thefnmark{}\@footnotetext}
\makeatother

\begin{document}

\graphicspath{{figure/}}
\selectlanguage{english}

\title{Exploiting Cellular Data for Disease Containment and \\Information Campaigns Strategies in Country-Wide Epidemics\footnote{Appeared in Proceedings of NetMob 2013. Boston, MA, USA. May 2013. \\}}

\author{A. Lima}
\affiliation{School of Computer Science, University of Birmingham, United Kingdom}
\author{M. De Domenico}
\affiliation{School of Computer Science, University of Birmingham, United Kingdom}
\author{V. Pejovic}
\affiliation{School of Computer Science, University of Birmingham, United Kingdom}
\author{M. Musolesi}
\affiliation{School of Computer Science, University of Birmingham, United Kingdom}

\begin{abstract}

Human mobility is one of the key factors at the basis of the spreading of diseases in a population. Containment strategies are usually devised on movement scenarios based on coarse-grained assumptions. Mobility phone data provide a unique opportunity for building models and defining strategies based on very precise information about the movement of people in a region or in a country. Another very important aspect is the underlying social structure of a population, which might play a fundamental role in devising information campaigns to promote vaccination and preventive measures, especially in countries with a strong family (or tribal) structure.

In this paper we analyze a large-scale dataset describing the mobility and the call patterns of a large number of individuals in Ivory Coast. We present a model that describes how diseases spread across the country by exploiting mobility patterns of people extracted from the available data. Then, we simulate several epidemics scenarios and we evaluate mechanisms to contain the epidemic spreading of diseases, based on the information about people mobility and social ties, also gathered from the phone call data. More specifically, we find that restricting mobility does not delay the occurrence of an endemic state and that an information campaign based on one-to-one phone conversations among members of social groups might be an effective countermeasure. 
\end{abstract}

\maketitle

\flushbottom


\input{introduction}
\input{dataset}

\input{model}
\input{analysis}
\input{conclusion}

\begin{acknowledgments}
The authors thank Charlotte Sophie Mayer for useful and fruitful discussions. This work was supported through the EPSRC Grant ``The Uncertainty of Identity: Linking Spatiotemporal Information Between Virtual and Real Worlds'' (EP/J005266/1).
\end{acknowledgments}

\newpage
\addcontentsline{toc}{section}{References} 
\bibliographystyle{mprsty} 
\bibliography{d4d}

\end{document}

%% file: introduction.tex
\section{Introduction}


Health and well-being of populations are heavily influenced by their behaviour. The impact of the habits and local customs, including patterns of interactions and mobility at urban and regional scales, on health issues is remarkable~\cite{helman2001culture}. The diffusion of mobile technology we are experiencing nowadays gives scholars an unprecedented opportunity to study massive data that describe human behavior~\cite{lazer2009life}. An increasing number of people carries smart mobile phones, equipped with many sensors and connected to the Internet, for the whole day~\cite{internetcomputing}. 
Data coming from a large number of people can describe trends in the macroscopic behavior of populations~\cite{onnela2007structure,gonzalez2008understanding,eagle2009eigenbehaviors}. The results of the analysis of these trends can be directly applied to a number of real-world scenarios, and, more in general, to several applications where cultural and local differences play a central role. Analyzing this kind of data can provide invaluable help to support the decision-making process, especially in critical situations. For this reason, many public and private organizations are nowadays increasingly adopting a data-centric approach in their decisional process~\cite{mckinsey}. We believe that this strategy can be particularly useful in developing countries, which might have a lacking infrastructure\footnote{We use the term ``developing'' to indicate countries that are assigned a low Human Development Index (HDI) by United Nations Statistics Division. We are aware of the limitations of this classification. As reported by UN, \textit{the designations ``developed'' and ``developing'' are intended for statistical convenience and do not necessarily express a judgment about the stage reached by a particular country or area in the development process}~\cite{unsd}.}.

Among the issues that developing countries are facing today, healthcare is probably the most urgent~\cite{healthpeople}. In these countries the effectiveness of campaigns is often reduced due to low availability of data, inherent limits in the infrastructure and difficult communication with the citizens, who might live in vast and remote rural areas. As a result, action plans are difficult to deliver. However, we believe that a data-centric approach can be an innovative and effective way to address these issues.

In this paper, we focus on containment of epidemics. We use movement data extracted from the registration patterns in a cellular network to evaluate the influence of human mobility on the spreading of diseases in a geographic area. In particular, we utilize this model to investigate how infectious agents might spread to distant locations because of human movement in order to identify optimal strategies that can be adopted to contrast the epidemics.
We also evaluate how the collaborative effort of the population can be crucial in critical scenarios. For the reasons we mentioned before, in countries that are facing development challenges, vaccination campaigns are often hard to advertise to the population. Lack of communication and information is believed to be among the main causes of failure for immunization campaigns. The same applies to awareness campaigns that try to promote prophylaxis procedures that reduce the occurrence of contagion. However, in these cases, we argue that a collaborative effort leveraging individual social ties can be effective in propagating effective information (i.e., a sort of ``immunizing information'') to a widespread audience. Moreover, information received by people who are socially close can have a higher chance of leading to an actual action.

A large body of research has been conducted on models that describe the diffusion of diseases, with a particular recent interest on the role that human movement plays in spreading infections in large geographic areas~\cite{colizza_modeling_2007,epstein_controlling_2007,meloni_modeling_2011}, and also on the impact of human behavior on the spreading itself~\cite{funk_modelling_2010,meloni_modeling_2011}. With respect to the state of the art, the main contributions of this paper can be summarized as follows:
\begin{itemize}
\item
We propose an epidemic model based on a network of geographic metapopulations, which describes how people move between different geographic regions and spread the disease. 
\item
We evaluate containment techniques based on the restriction of mobility of the most central areas. The centrality of the areas is extracted by building a movement network between all the geographic areas based on the mobility patterns of the individuals.
\item
We extend the model with a competing information spreading where \textit{distance contagion} might take place. In other words, we study the dynamics of the system considering three characterizing aspects of the problem: the disease epidemics, human mobility and information spreading. This epidemics represents the diffusion of information related to measures to prevent or to combat the diseases, such as information about the ongoing vaccination and prevention campaigns in a certain area or actions that will help to limit spread of the infection, such as boiling water or avoiding contacts with people that are already ill. 
\item We evaluate the models by using the data provided by the Orange ``Data for Development''~\cite{blondel2012data}. We discuss the effectiveness of the containment strategies and, in particular, for the information dissemination strategy, we identify the degree of participation that is required to make it successful. 
\item We observe that restricting mobility by disallowing any movement from and to a limited set of subprefectures does not delay the occurrence of the endemic state in the rest of the country. We also find that a collaborative effort of prevention information spreading can be an effective countermeasure.
\end{itemize}

This paper is organized as follows. In Sec.\,\ref{sec:data} we briefly describe the four different datasets provided by Orange and we specify how they are used in the present study. In Sec.\,\ref{sec:model} we introduce our two models for epidemics and information spreading by taking into account human mobility and call patterns observed in Ivory Coast. In Sec.\,\ref{sec:analysis} we present the results obtained by simulating several epidemics scenarios and evaluating mechanisms to contain the epidemic spreading of diseases. Finally, in Sec.\,\ref{sec:conclusion} we summarize our main findings and we propose how the present study can be improved if more detailed data about mobility and calls will be available.

%% file: dataset.tex
\section{Overview of the Dataset}\label{sec:data}

The data provided for the D4D challenge~\cite{blondel2012data} consist of four datasets (identified by the labels SET1, SET2, SET3, SET4), containing information about user mobility and call patterns at various levels of granularity and time duration. We will now discuss how these datasets can be used to build a model which accounts for user mobility and information spreading.

\begin{figure}[t]
\center
 \subfigure[]
   {
   \includegraphics[width=.45\linewidth]{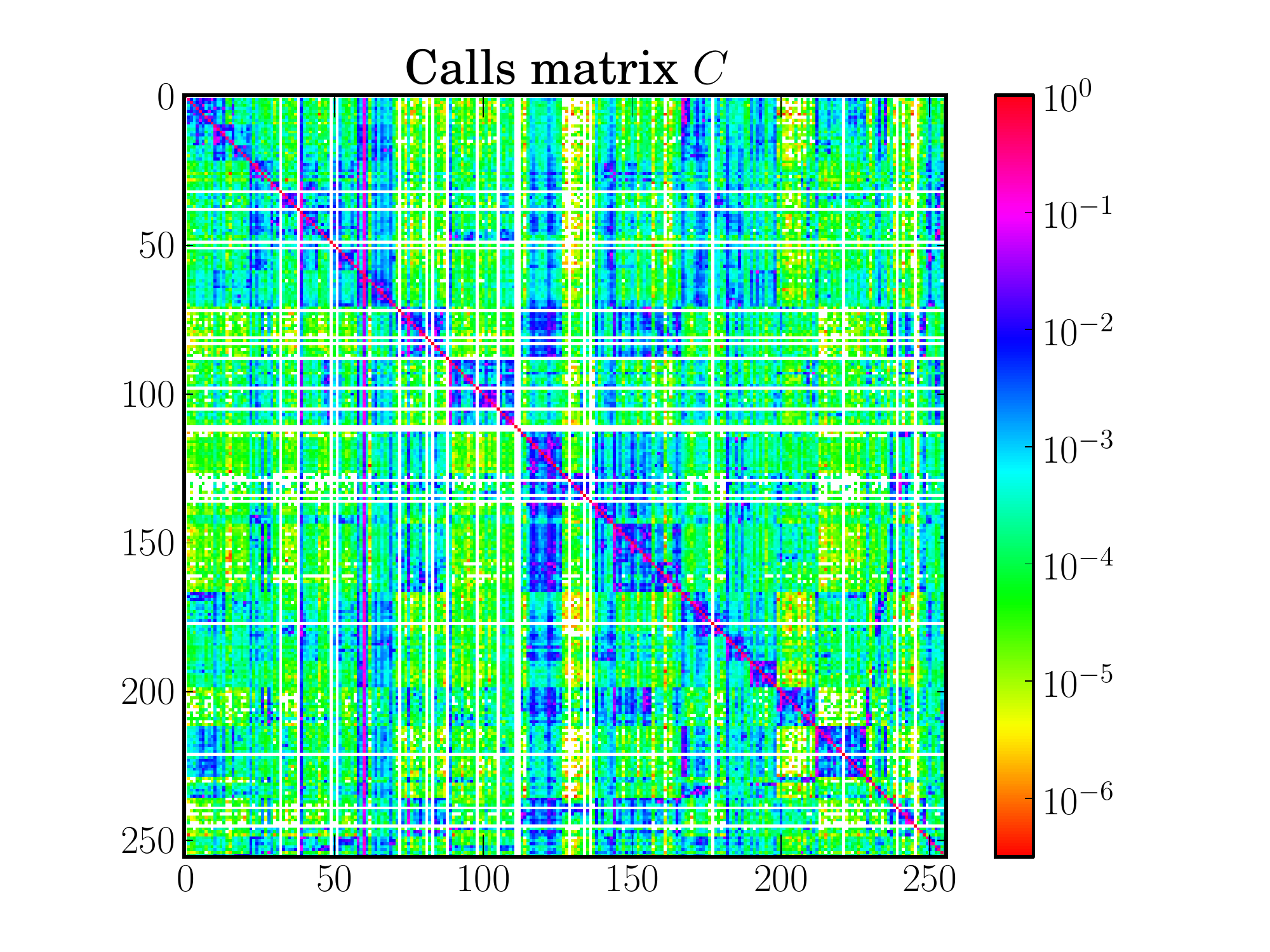}
   \label{fig:calls-matrix}
   }
\subfigure[]
   {
   \includegraphics[width=.45\linewidth]{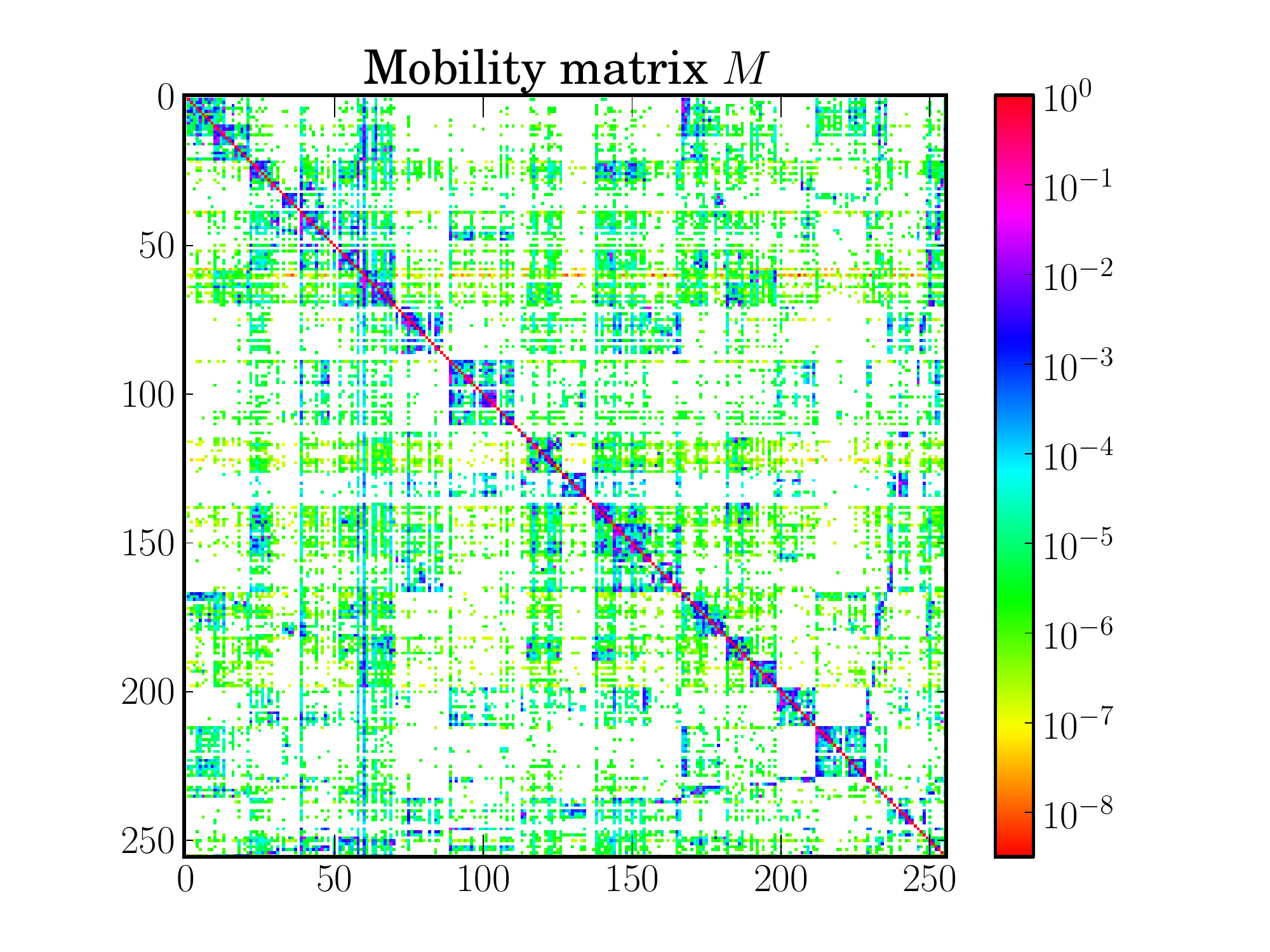}
   \label{fig:mobility-matrix}
   }
 \caption{Logarithmic representation of the calls matrix~(a) and the mobility matrix (b). Null values are indicated using the white color. For both matrices highest values are mostly concentrated along the diagonal, showing that communication and movement between sub-prefectures is highly uncommon. However, the calls matrix is visibly denser than the mobility matrix, confirming that phone contacts between different sub-prefectures are more usual than movement.}
\end{figure}

\begin{figure*}[!t]
\subfigure[]
  {
  \includegraphics[width=.35\linewidth]{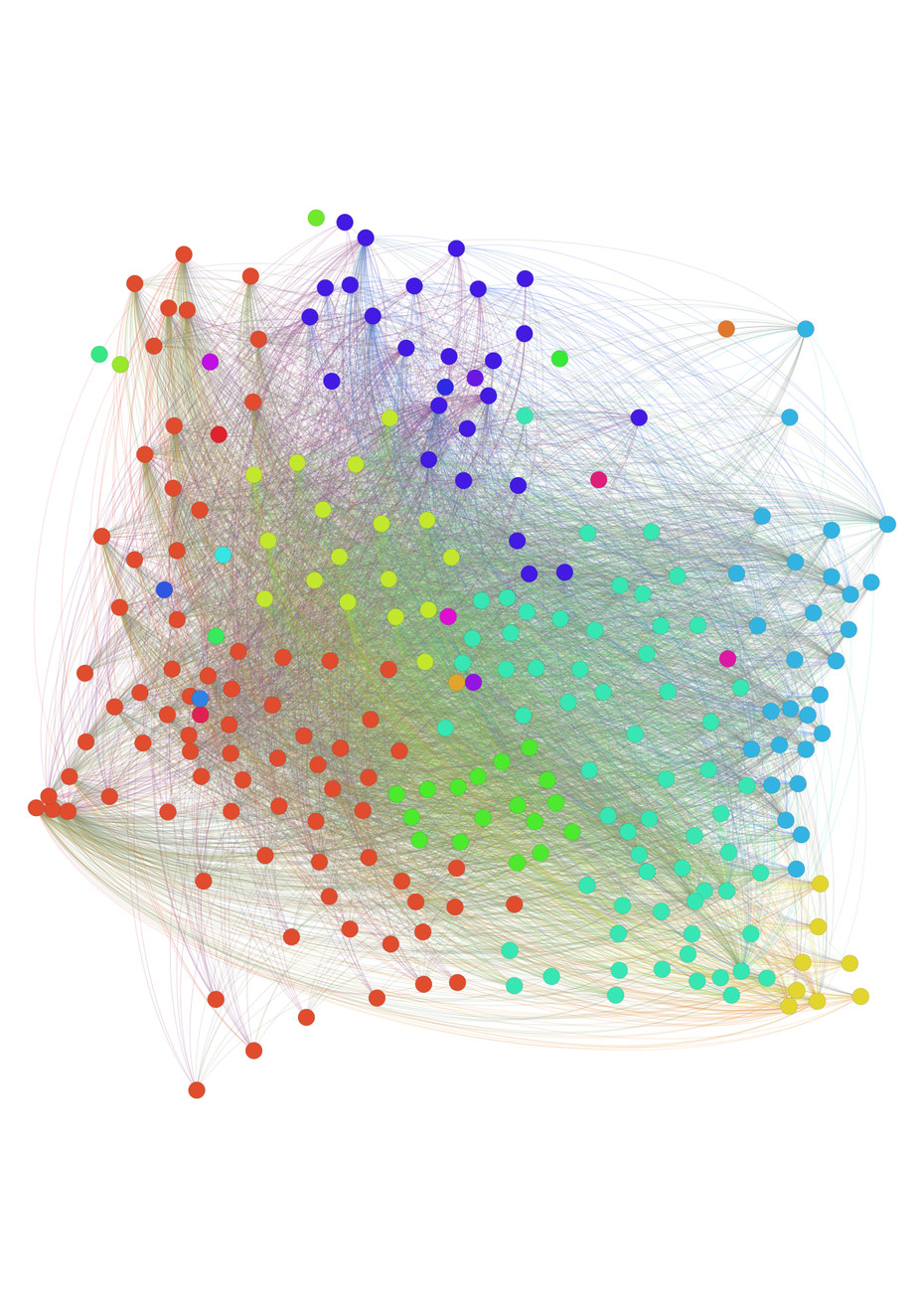}
  \label{fig:calls-modul}
  }
\hspace{5mm}
\subfigure[]
  {
  \includegraphics[width=.35\linewidth]{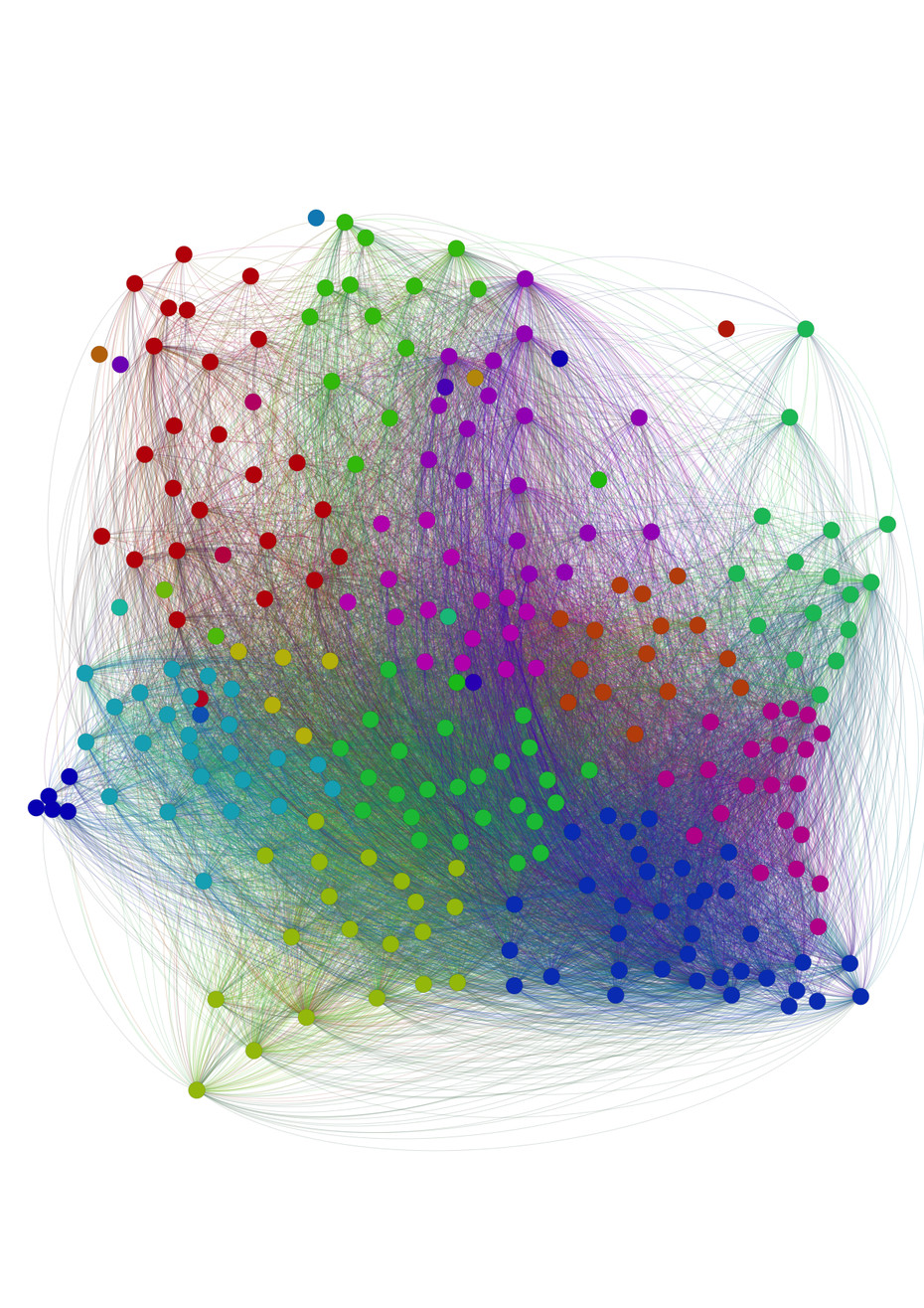}
  \label{fig:mobility-modul}
  }
\caption{Geographic network obtained from call logs (a) and mobility traces (b). Color is used to indicate the community structure: nodes within the same community are represented with the same color.}
\end{figure*}

Two datasets contain information about mobility and communication patterns at macroscopic level. More precisely:
\begin{itemize}

\item
The SET1 dataset contains the number and the duration of calls between pairs of cell phone towers, aggregated by hour. This dataset provides macroscopic information about communication in the country. We associate cell phone towers with the sub-prefecture they are located in, by using the supplied geographic position. Then, we evaluate the probability of a call being established between sub-prefectures $i$ and $j$ with:
\begin{eqnarray}
\label{def:cmatrix}
c_{ij} = \frac{\mathcal{C}_{ij}}{\sum_k \mathcal{C}_{ik}},
\end{eqnarray}
where $\mathcal{C}_{ij}$ is the number of phone calls initiated from the sub-prefecture $i$ and directed to the sub-prefecture $j$, during the entire period of observation. The term at denominator indicates the total communication flux between every pair of sub-prefectures and it is used to normalize the probability. Using these values we build a calls matrix $C$, shown in Fig.~\ref{fig:calls-matrix}. This matrix also shows high values along the diagonal, but it is distinctly denser, showing that calls between sub-prefectures are more common than movement. The vertical line at $x=60$ identifies calls directed to the sub-prefecture that contains the capital.

\item
The SET3 dataset contains the trajectories of 50,000 randomly-selected individuals, at a sub-prefecture level resolution, for five months.\footnote{17 sub-prefectures do not have any cell phone towers and for this reason do not appear in SET3. We discard these sub-prefectures from our analysis, since their users will be considered as belonging to nearby sub-prefectures.} This dataset can be used to estimate the probability that an individual moves from the sub-prefecture $i$ to the sub-prefecture $j$:
\begin{equation}
\label{def:mmatrix}
m_{ij} = \frac {\sum_u \mathcal{M}_{ij}^u} {\sum_k \sum_u \mathcal{M}_{ik}^u},
\end{equation}
where $\mathcal{M}_{ij}^u$ is the number of times user $u$ moves from the sub-prefecture $i$ to $j$. The numerator counts how many times users who are in $i$ move to $j$; the denominator normalizes this number by the total number of transitions from $i$ to any sub-prefecture $k$. Using these values we build a mobility matrix $M$, shown in Fig.~\ref{fig:mobility-matrix}. By using this matrix, we model human mobility in the country as a Markov process~\cite{norris1998markov}. We observe that the matrix is quite sparse and the highest values are concentrated along the diagonal. As the representation is in logarithmic scale, this demonstrates that the movement between sub-prefectures is present, but rather uncommon. 

\end{itemize}

In Fig.\,\ref{fig:calls-modul} and Fig.\,\ref{fig:mobility-modul} we show the geographic networks of calls and mobility, respectively. Nodes are positioned using the geographic locations of the sub-prefecture they represent, and their color indicates the community structure of the network based on~\cite{blondel2008fast}.

The other two datasets provide microscopic information about mobility and communication patterns between individuals. Although we do not use them for the analysis in this paper, we now briefly outline how they could be used:

\begin{itemize}
\item
The SET2 dataset contains fine-grained individual trajectories of 50,000 randomly sampled individuals over two-week periods. This dataset could be used to estimate the number of potential connections that an individual might have in a certain area, served by a cell phone tower.

\item
The SET4 dataset contains time-varying ego-networks of 5,000 users, describing the network of communication in time-slots of 2 weeks. If two users are connected by a link in a time-slot, it means that \textit{at least} one call occurred during the two weeks under consideration\footnote{We have found that 1.31\% of the total number of edges in ego-networks connect pairs of users who are neither egos nor first-level neighbors: therefore, we do not consider such edges in our analysis.}. The ego-network aggregated over the whole observation time, built considering every link that is present at least once, describes the number of people contacted by an individual during the entire period. This dataset could be used to estimate the number of potential social connections that an individual might get in touch with. The degree distribution of the aggregate ego-network is shown in Fig.~\ref{fig:ego_friends}.

\begin{figure}[t]
   \includegraphics[width=8.cm]{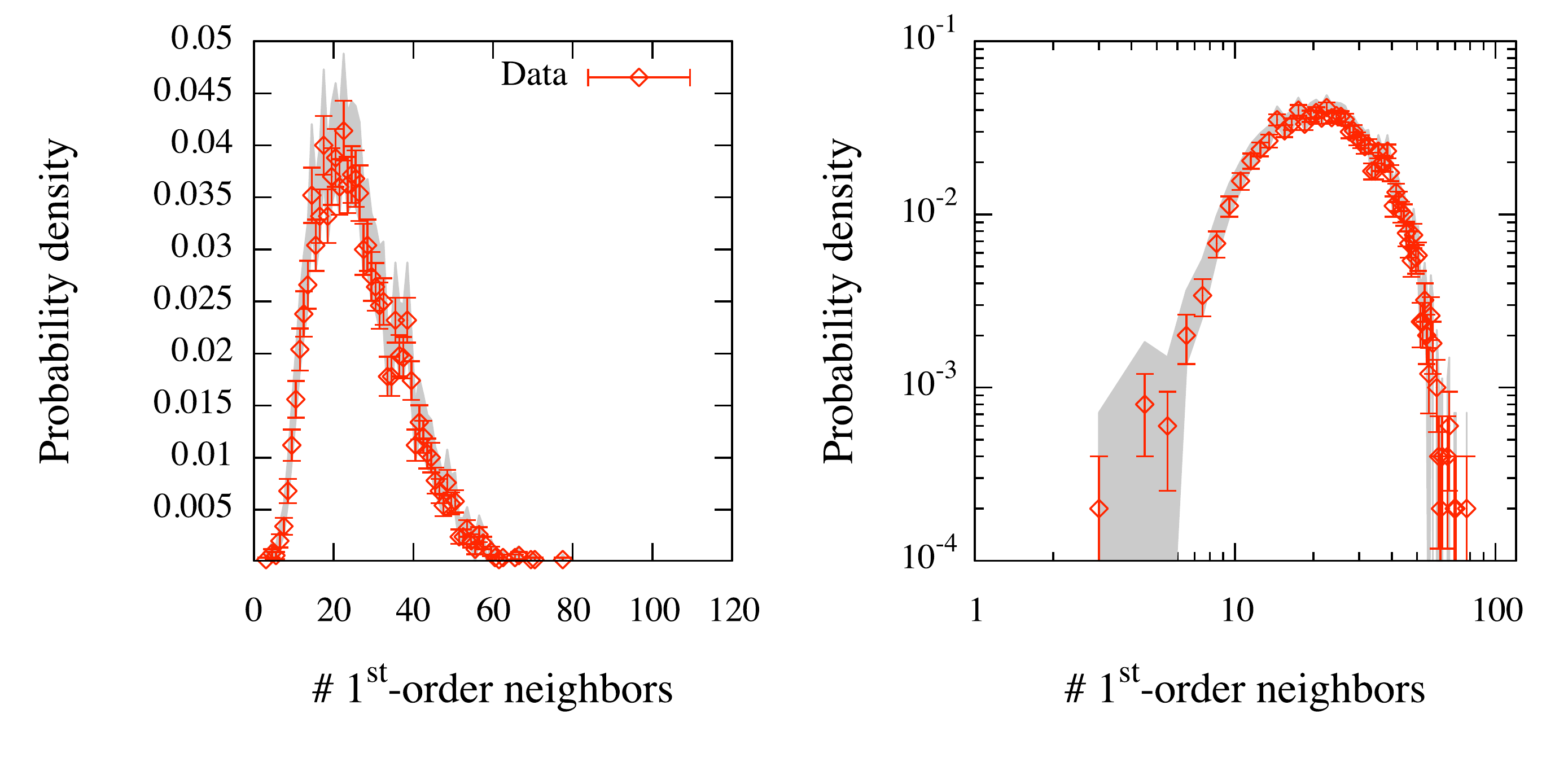}
   \caption{Distribution of friends for the ego-networks aggregated over time. Error bars indicate statistical uncertainty, while the shaded area represent 99\% confidence intervals around the observed value.}
   \label{fig:ego_friends}
\end{figure}

\end{itemize}

%% file: model.tex
\section{Spreading Models}\label{sec:model}

In this section we discuss two models: a model of disease spreading as a function of the mobility patterns of individuals between different geographic areas inferred from the cellular registration records and a model for information spreading among the same population, considering the social structure inferred from the call records. In the following section, we will evaluate the models using the data provided for the Orange Data for Development challenge.

\subsection{Epidemic Spreading and Mobility}

We will now present a model that represents the evolution of an epidemic taking place on a network of metapopulations. The aim of the model is to describe how the system evolves under the action of two processes, contagion and mobility. For this dataset, each metapopulation is composed by the individuals located in a particular sub-prefecture. Hence, the population is distributed in $n$ different metapopulations, each having $N_i[t]$ individuals at time $t$. We make the simplifying assumption that there are no deaths and births in the considered time window, i.e., at each time $t = 1, 2, \dots T$ the total population is constant $\sum_{i=1}^n N_i[t] = N$.

We assume that contagion happens inside each metapopulation following a standard SIS model~\cite{keeling2011modeling}. We indicate the number of infected and susceptible individuals at time $t$ in a sub-prefecture $i$ with $I_{i}[t]$ and $S_{i}[t]$, respectively. At each time $t$ a person is either infected or susceptible, therefore $N_{i}[t] = I_{i}[t] + S_{i}[t]$.

Simultaneously, individuals move through the metapopulation network according to the \textit{mobility matrix} $M$ of dimension $n \times n$ extracted from the cellular traces. The generic element $m_{ij}$ of the matrix represents the probability that a person moves from the metapopulation $i$ to $j$, as described by Eq.\,\ref{def:mmatrix}\footnote{In general, this matrix can be time-varying, and it can be adjusted according to seasonal trends or real-time data at each step, for example following estimates based on historical data. In particular, this matrix can be used to study the impact of policies in real-time. However, in order to simplify the presentation, we use a matrix not changing over time. The treatment can be generalized, also applying the recent theoretical results related to time-varying networks~\cite{TSMML10:smallworld,holme2012temporal}.}. This matrix describes how the state variables $N_i[t]$ evolve over time: $N_i[t+1] = \sum_{j=1}^n m_{ji} N_i[t]$. Under the assumption that individuals inside the classes $I$ and $S$ move consistently we can write the last relation also for the state variables $I_i[t]$ and $S_i[t]$\footnote{This assumption can also be relaxed when data about the different classes of individuals is available, i.e., when a matrix for each class can be defined.}. The contagion-mobility combined system can then be described by the following set of equations:

\begin{eqnarray}
I_i[t+1] &=& \sum_{j=1}^n m_{ji} \left[ I_j[t] + \lambda\frac{S_j[t]}{N_j[t]} I_j[t] - \gamma I_j[t] \right] \nonumber \\
S_i[t+1] &=& \sum_{j=1}^n m_{ji} \left[ S_j[t] - \lambda\frac{S_j[t]}{N_j[t]} I_j[t] + \gamma I_j[t] \right] \nonumber ,
\end{eqnarray}

for each sub-prefecture $i = 1, 2, \dots, n$, with $\lambda$ being the product of contact rate and contagion probability and $\gamma$ being the recovery rate. The formulae inside the square brackets describe the evolution of $n$ SIS models, one for each metapopulation. They are multiplied for the elements of the mobility matrix, which accounts for individuals moving between metapopulations.

This analytical model describes the expected outcome of a stochastic model where the following actions occur at each time step:
\begin{enumerate}
\item Each infected person in the sub-prefecture $j$ causes the infection of new $\lambda\frac{S_j}{N_j}$ individuals inside $j$. This step is repeated for each sub-prefecture.
\item A new position $i$ is assigned to each individual in the sub-prefecture $j$ according to the probability density function $[m_{j1},m_{j2},\dots,m_{jn}]$. This step is repeated for each sub-prefecture.
\end{enumerate}
%
%
\subsection{Information Spreading}
\begin{figure}[t!]
\center
\begin{tikzpicture}[>=stealth',shorten >=1pt,auto,node distance=2cm]
  \node[state] (R)      {$R$};
  \node[state,initial above]         (S) [right of=R]  {$S$};
  \node[state]         (I) [right of=S] {$I$};
 
  \path[<-]  (R)   edge [bend left]  node {$\omega$} (S);
  \path[<-]  (S)   edge [bend left]  node {$\gamma$} (I);
  \path[<-]  (I)   edge [bend left]  node {$\lambda$} (S);
  \path[<-]  (S)   edge [bend left]  node {$\xi$} (R);
\end{tikzpicture} \\
\begin{tikzpicture}[>=stealth',shorten >=1pt,auto,node distance=2cm]
  \node[state] (U) [initial left]  {$U$};
  \node[state] (A) [accepting, right of=U]                  {$A$};
 
  \path[<-]    (A)   edge [bend left]  node {$\psi$} (U);
\end{tikzpicture}
\caption{State machines describing the state transitions of a person with respect
to the disease contagion (R=Resistant, S=Susceptible and I=Infected) and with respect to
the information spreading (U=unaware, A=aware), respectively. A
person starts in the susceptible and unaware states. We assume that aware individuals spread the information and cannot go back to the unaware state.}
\label{fig:sm}
\end{figure}
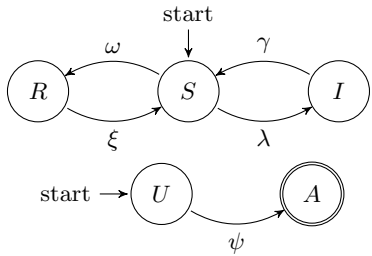
%
%
The model we presented in the last section tries to reproduce the spreading of a disease in a population where individuals change locations over time. The aim of this work is to analyze some scenarios and study the effectiveness of some containment techniques. In particular, as anticipated, we would like to investigate if a collaborative effort of the population is able, in theory, to reduce considerably the spread of the disease and what proportions should it have to be effective. More precisely, the population can disseminate information through personal social ties \textit{immunizing}, such as information about prevention techniques, hygiene practises, advertisement of nearby vaccination campaigns and in general any information that can lead to a reduction of the number of contagion events.

In order to take into consideration these aspects, we now use a SIR model for each metapopulation, so that each person either belongs to the susceptible (S), infected (I) or resistant (R) category. At the same time, another simultaneous epidemic happens on the network of metapopulations, disseminating information that can make individuals resistant to the disease. In fact, a person also belongs to the category of unaware (U) or aware (A) individuals, with respect to the immunizing information. More formally, we have that $N_i[t] = I_i[t] + S_i[t] + R_i[t] = A_i[t] + U_i[t]$. 

It is worth noting that this ``immunizing epidemic'' goes \textit{beyond} the boundaries of metapopulations (sub-prefectures): in other words, it is a \textit{distance contagion}. It is also important to remark that the states ``aware'' and ``resistant'' are substantially different. An unaware person that receives the information (i.e. has an ``information contact'') becomes aware with rate $\psi$; since the person is aware, he or she will start spreading the information as well. An infected person that receives the information becomes immune with rate $\omega$. Additionally, individuals who have acquired immunity through information can lose it with rate $\xi$. The transition rates between states are summarized in Fig.~\ref{fig:sm}.
The model can be described by the following set of equations, specifying how state vectors evolve over time:

\begin{alignat}{2}
I_i[t+1] = {}& \sum_{j=1}^n m_{ji} {} \bigg[  I_j[t] && + \lambda\frac{S_j[t]}{N_j[t]} I_j[t] - \gamma I_j[t] \bigg] \nonumber \\
S_i[t+1] = {}& \sum_{j=1}^n m_{ji} {} \bigg[  S_j[t] && - \lambda\frac{S_j[t]}{N_j[t]} I_j[t] + \gamma I_j[t] + \xi R_j[t] + \nonumber \\
           {} & {} &&- \omega S_j[t] \frac{\sum_{k=1}^n {c_{kj}{A_k[t]}}}{\sum_{k=1}^n c_{kj}N_k[t]} \bigg] \nonumber\\
R_i[t+1] = {}& \sum_{j=1}^n m_{ji} \bigg[ R_j[t] && - \xi R_j[t] + \omega S_j[t] \frac{\sum_{k=1}^n {c_{kj}{A_k[t]}}}{\sum_{k=1}^nc_{kj}N_k[t]} \bigg] \nonumber\\
A_i[t+1] = {}& \sum_{j=1}^n m_{ji} \bigg[ A_j[t] && + \psi U_j[t] \frac{\sum_{k=1}^n {c_{kj}{A_k[t]}}}{\sum_{k=1}^nc_{kj}N_k[t]} \bigg] \nonumber\\
U_i[t+1] = {}& \sum_{j=1}^n m_{ji} \bigg[ U_j[t] && - \psi U_j[t] \frac{\sum_{k=1}^n {c_{kj}{A_k[t]}}}{\sum_{k=1}^nc_{kj}N_k[t]} \bigg] \label{eq:model}
\end{alignat}

for every $i = 1, 2, \dots, n$. The fraction $\frac{\sum_{k=1}^{n} c_{kj}A_k[t]}{\sum_{k=1}^{n} c_{kj}N_k[t]}$ represents the probability that a call from an aware person occurs in the metapopulation $j$. It models the distance-contagion, and it is possible to verify that if the matrix is identical (absence of contacts between populations) it reduces to $A_k[t] / N_k[t]$, falling back to a model where contagion occurs only inside metapopulations.

This analytical model describes the expected value of a stochastic model where the following actions occur at each time step $t$:
\begin{enumerate}
\item Each infected person in the sub-prefecture $j$ causes $\lambda\frac{S_j}{N_j}$ new individuals to get infected, inside $j$. This step is repeated for each sub-prefecture.
\item Each unaware person in the sub-prefecture $j$ becomes aware with probability $\psi\frac{\sum_{k=1}^n {c_{kj}{A_k[t]}}}{\sum_{k=1}^nc_{kj}N_k[t]}$. This step is repeated for each sub-prefecture.
\item Each person in the sub-prefecture $j$ who is susceptible, becomes resistant with probability $\omega\frac{\sum_{k=1}^n {c_{kj}{A_k[t]}}}{\sum_{k=1}^nc_{kj}N_k[t]}$. This step is reapeated for each sub-prefecture.
\item A new position $i$ is assigned to each person in the sub-prefecture $j$ according to the probability density function $[m_{j1}, m_{j2}, \dots, m_{jn}]$. This step is repeated for each sub-prefecture.
\end{enumerate}

%% file: analysis.tex
\section{Analysis}\label{sec:analysis}


We initialize each scenario by allocating 22 million individuals (the estimated population size of Ivory Coast for July 2012 is 21,952,093~\cite{CIAwfact2012}) to different sub-prefectures across the country, according to the data in SET3. In each scenario we bootstrap the spreading process by infecting a fraction of the population ($0.1\%$) distributed across metapopulations according to different criteria:
\begin{itemize}
\item Uniform distribution: every sub-prefecture gets a number of infected proportional to their population, i.e., every sub-prefecture has the same fraction of infected population.
\item Random: a single sub-prefecture, chosen randomly, is the origin of the infection.
\item Centrality based: the sub-prefectures are ordered by decreasing centrality values, then the first 1, 5 or 10 highest ranked sub-prefectures are chosen, as shown in Table~\ref{table:centralities}.
\end{itemize}

\begin{table}[b]
    \begin{tabular}{|c|c|c|c|}
        \hline
        Betweeness & Closeness & Degree & Eigenvalue \\ \hline
        60  & 60  & 60  & 60  \\
        39  & 58  & 58  & 58  \\ 
        89  & 39  & 39  & 39  \\ 
        58  & 69  & 69  & 69  \\ 
        75  & 138 & 138 & 250 \\ 
        144 & 250 & 64  & 138 \\ 
        138 & 64  & 144 & 64  \\ 
        165 & 144 & 250 & 144 \\ 
        212 & 182 & 122 & 122 \\ 
        168 & 122 & 182 & 182 \\
        \hline
    \end{tabular}
    \caption{Highest ranked sub-prefectures, according to different definitions of centrality. We observe that the sets of the top 10 sub-prefectures ordered by centrality are very similar.}
    \label{table:centralities}
\end{table}

We study the evolution of the epidemics for a period of 6 months. We investigate multiple scenarios using the analytical model considering a large set of ranges for the key parameters. We conducted a series of Monte-Carlo simulations for multiple sets of parameters, confirming the validity of the analytical models presented in the previous section. In the following, we present results based on these models.

\subsection{No Countermeasures}

\begin{figure}[!t]
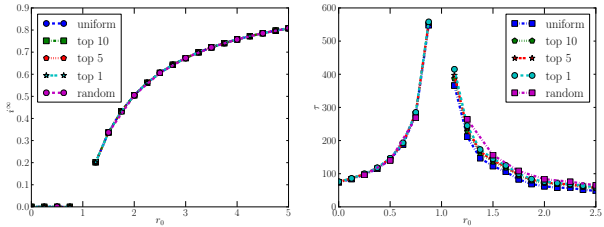

\center
\includegraphics[width=.46\linewidth,trim=10mm 0 10mm 0]{{{figure/iinf_r0_nocalls}}}
\includegraphics[width=.46\linewidth,trim=10mm 0 10mm 0]{{{figure/tau_r0_nocalls}}}
 \caption{Fraction of infected population at the stationary state (left panel) and time required to reach the stationary state (right panel), for different values of $r_0$ and for different initial conditions. Missing values in the curves mean that, for the corresponding values, no stationary state is reached during the period of observation.}
 \label{fig:iinf-tau-nocalls}
\end{figure}

We will firstly explore the evolution of the epidemics in the case where no countermeasures are taken. 
In order to analyze the evolution of the system more clearly, we investigate two measures: the fraction of infected population $i^\infty$ at the stationary state and the time required to reach the stationary state $\tau$. In Fig.~\ref{fig:iinf-tau-nocalls} we plot their values versus $r_0 = \frac{\lambda}{\gamma}$, which is the basic reproductive ratio of a classic SIS model~\cite{keeling2011modeling}. As a future work, we plan to derive the analytical form of the basic reproductive ratio of our models, which take into account mobility and information spreading. We observe that for $r_0 = \frac{\lambda}{\gamma} < 1$ there is no endemic state (i.e., the final fraction of infected population is zero), whereas for $r_0 > 1$ a non-null fraction of population is infected. Values for $r_0=1$ are missing since no stationary state is reached within our observation window. In other words, for this particular scenario, experimental results show that the basic reproductive ratio of our model is very close to $r_0$; we expect this to be a consequence of the low inter-subprefectures mobility. We can also notice that the initial conditions do not affect  $i^\infty$ at all. Before the critical point (i.e., $r_0=1$) the choice of the initial conditions has also no impact on the delay time, whereas for $r_0 > 1$ it slightly affects the delay: epidemics that initially involves more sub-prefectures are slightly faster than the others.

\subsection{Geographic Quarantine}

\begin{figure}[t]
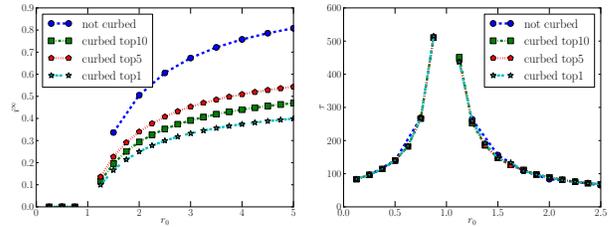

\center
\includegraphics[width=.46\linewidth,trim=10mm 0 10mm 0]{{{figure/iinf_r0_nocalls_curbed}}}
\includegraphics[width=.46\linewidth,trim=10mm 0 10mm 0]{{{figure/tau_r0_nocalls_curbed}}}
 \caption{Fraction of infected population at the stationary state (left panel) and time required to reach the stationary state (right panel), for different values of $r_0$ when the epidemic starts from a random sub-prefecture, and different levels of geographic quarantine are applied. Missing values in the curves mean that, for the corresponding values, no stationary state is reached during the period of observation.}
 \label{fig:iinf-tau-nocalls-curbed}
\end{figure}

\begin{figure*}[t]
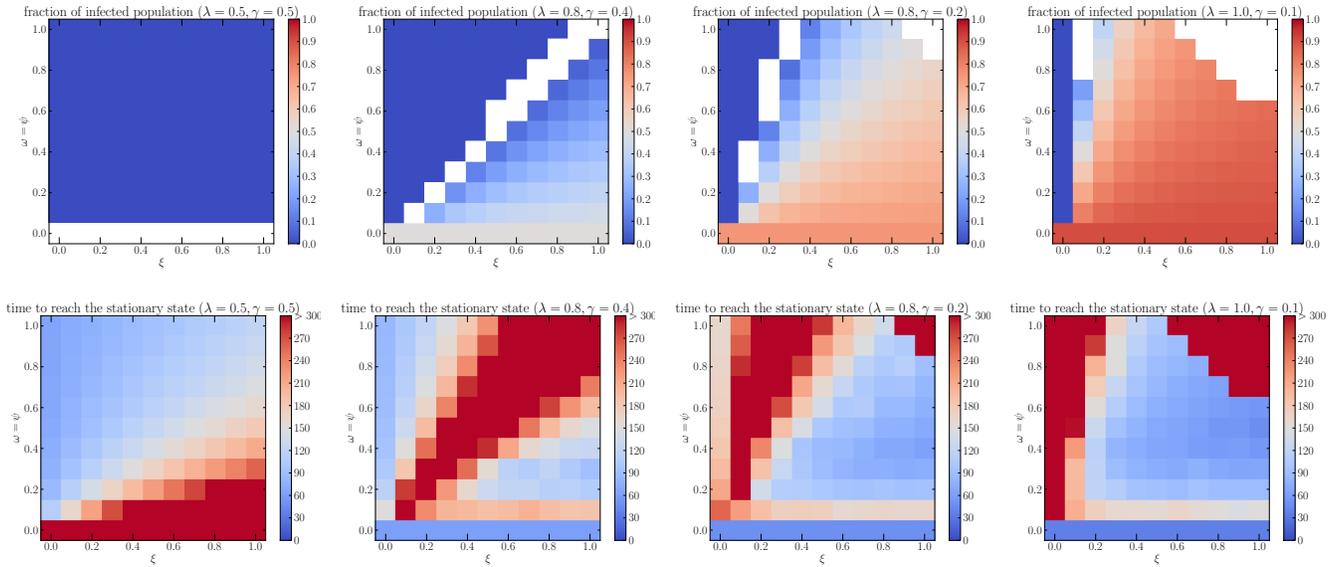

\center
\subfigure
   {
   \includegraphics[width=.23\linewidth,trim=20mm 0 15mm 0]{{{figure/iinf_heatmap_call1_l0.5_g0.5}}}
   }
   {
   \includegraphics[width=.23\linewidth,trim=20mm 0 15mm 0]{{{figure/iinf_heatmap_call1_l0.8_g0.4}}}
   }
   {
   \includegraphics[width=.23\linewidth,trim=20mm 0 15mm 0]{{{figure/iinf_heatmap_call1_l0.8_g0.2}}}
   }
   {
   \includegraphics[width=.23\linewidth,trim=20mm 0 15mm 0]{{{figure/iinf_heatmap_call1_l1.0_g0.1}}}
   }\\
   {
   \includegraphics[width=.23\linewidth,trim=20mm 0 15mm 0]{{{figure/tau_heatmap_call1_l0.5_g0.5}}}
   }
   {
   \includegraphics[width=.23\linewidth,trim=20mm 0 15mm 0]{{{figure/tau_heatmap_call1_l0.8_g0.4}}}
   }
   {
   \includegraphics[width=.23\linewidth,trim=20mm 0 15mm 0]{{{figure/tau_heatmap_call1_l0.8_g0.2}}}
   }
   {
   \includegraphics[width=.23\linewidth,trim=20mm 0 15mm 0]{{{figure/tau_heatmap_call1_l1.0_g0.1}}}
   }
 \caption{Fraction of infected population at the stationary state (top row) and time required to reach the stationary state (bottom row), for different values of $r_0 = \frac{\lambda}{\gamma}$ (from left to right 1, 2, 4, 10, respectively). White spaces show that no stationary state is reached during the period of observation.}
 \label{fig:heatmap-wP-x}
\end{figure*} 

We now analyze the effects of curbing on the mobility between sub-prefectures, i.e., forbidding all the incoming and outgoing movement of a group of sub-prefectures. In order to do so, we calculate the centrality values of each sub-prefecture in the mobility matrix. We present the results for eigenvalues centrality. As it is possible to observe in Tab.~\ref{table:centralities}, the ranking based on other centralities is very similar. Then, for the quarantine operations, we select those with the highest centrality values. From a practical point of view, this is achieved by simply changing the $i-$th row and column in the mobility matrix, so that all the elements $m_{ij}$ and $m_{ji}$ are null, except for the elements $m_{ii}=1$. For these scenarios, we randomly choose a single sub-prefecture where the initial individuals are infected, and then we average $i^\infty$ and $\tau$ over all runs. As shown in Fig.~\ref{fig:iinf-tau-nocalls-curbed}, the fraction of the infected population is sensibly affected by this measure, as the population inside the quarantined areas is protected from contagion. However, contrary to the intuition, the delay is not affected by the quarantine, even when the countermeasures involves 10 sub-prefectures, which account for almost half population. This suggests that such an invasive, expensive and hard to enforce measure reduces considerably the endemic size, but does not slow down the disease spreading in the rest of the country. For this reason, we now investigate a radically different approach to protect the population.

\begin{figure*}[t]
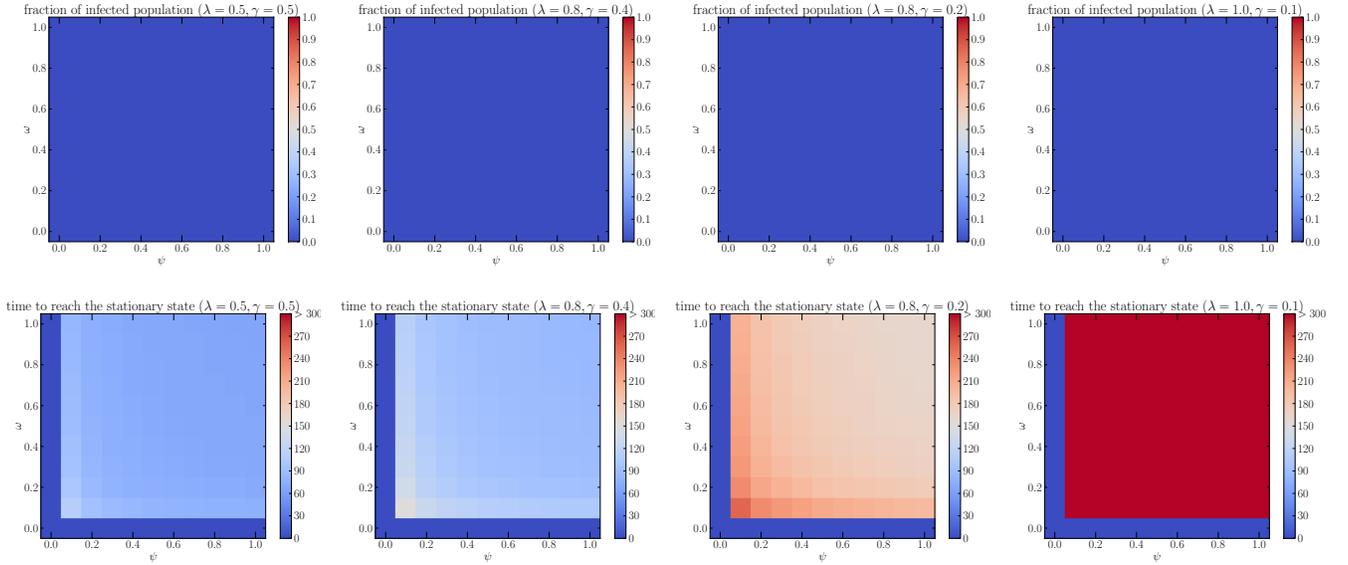

\center
\subfigure
   {
   \includegraphics[width=.23\linewidth,trim=20mm 0 15mm 0]{{{figure/iinf_heatmap_call1_l0.5_g0.5_xi0.0}}}
   }
   {
   \includegraphics[width=.23\linewidth,trim=20mm 0 15mm 0]{{{figure/iinf_heatmap_call1_l0.8_g0.4_xi0.0}}}
   }
   {
   \includegraphics[width=.23\linewidth,trim=20mm 0 15mm 0]{{{figure/iinf_heatmap_call1_l0.8_g0.2_xi0.0}}}
   }
   {
   \includegraphics[width=.23\linewidth,trim=20mm 0 15mm 0]{{{figure/iinf_heatmap_call1_l1.0_g0.1_xi0.0}}}
   }\\
   {
   \includegraphics[width=.23\linewidth,trim=20mm 0 15mm 0]{{{figure/tau_heatmap_call1_l0.5_g0.5_xi0.0}}}
   }
   {
   \includegraphics[width=.23\linewidth,trim=20mm 0 15mm 0]{{{figure/tau_heatmap_call1_l0.8_g0.4_xi0.0}}}
   }
   {
   \includegraphics[width=.23\linewidth,trim=20mm 0 15mm 0]{{{figure/tau_heatmap_call1_l0.8_g0.2_xi0.0}}}
   }
   {
   \includegraphics[width=.23\linewidth,trim=20mm 0 15mm 0]{{{figure/tau_heatmap_call1_l1.0_g0.1_xi0.0}}}
   }
 \caption{Fraction of infected population at the stationary state (top row) and time required to reach the stationary state (bottom row), for different combinations of $\frac{\lambda}{\gamma}$ (from left to right 1, 2, 4, 10, respectively) and $\xi=0$. White spaces show that no stationary state is reached during the period of observation.}
 \label{fig:heatmap-w-P-x0.0}
\end{figure*} 

\begin{figure*}[t]
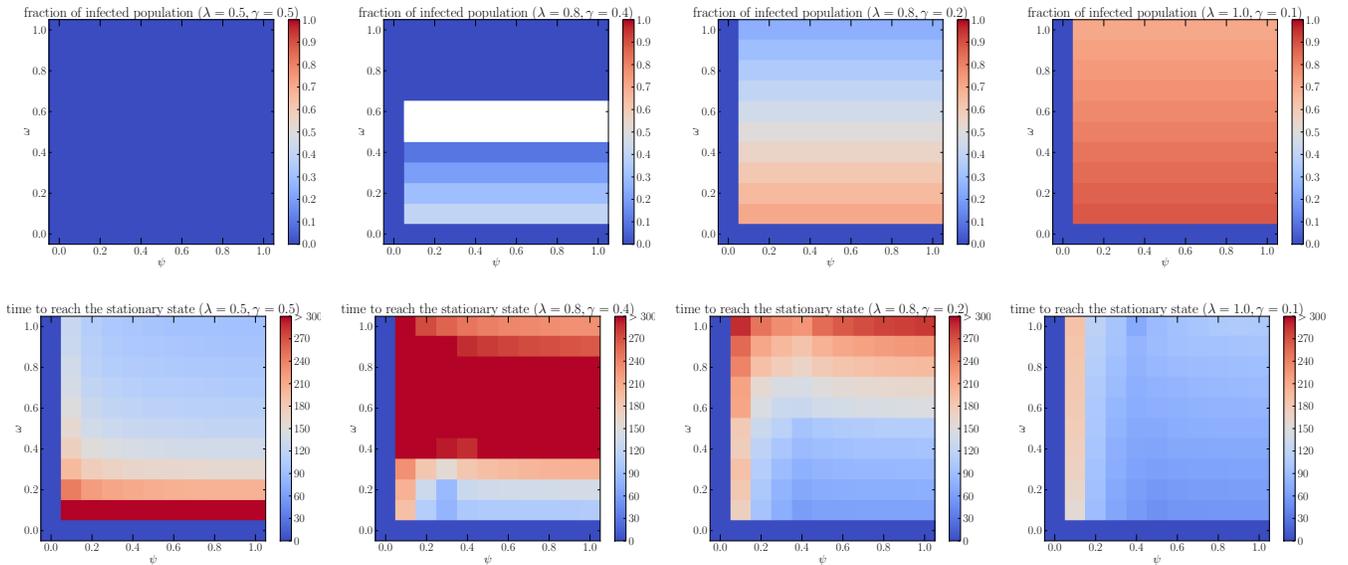

\center
\subfigure
   {
   \includegraphics[width=.23\linewidth,trim=20mm 0 15mm 0]{{{figure/iinf_heatmap_call1_l0.5_g0.5_xi0.5}}}
   }
   {
   \includegraphics[width=.23\linewidth,trim=20mm 0 15mm 0]{{{figure/iinf_heatmap_call1_l0.8_g0.4_xi0.5}}}
   }
   {
   \includegraphics[width=.23\linewidth,trim=20mm 0 15mm 0]{{{figure/iinf_heatmap_call1_l0.8_g0.2_xi0.5}}}
   }
   {
   \includegraphics[width=.23\linewidth,trim=20mm 0 15mm 0]{{{figure/iinf_heatmap_call1_l1.0_g0.1_xi0.5}}}
   }\\
   {
   \includegraphics[width=.23\linewidth,trim=20mm 0 15mm 0]{{{figure/tau_heatmap_call1_l0.5_g0.5_xi0.5}}}
   }
   {
   \includegraphics[width=.23\linewidth,trim=20mm 0 15mm 0]{{{figure/tau_heatmap_call1_l0.8_g0.4_xi0.5}}}
   }
   {
   \includegraphics[width=.23\linewidth,trim=20mm 0 15mm 0]{{{figure/tau_heatmap_call1_l0.8_g0.2_xi0.5}}}
   }
   {
   \includegraphics[width=.23\linewidth,trim=20mm 0 15mm 0]{{{figure/tau_heatmap_call1_l1.0_g0.1_xi0.5}}}
   }
 \caption{Fraction of infected population at the stationary state (top row) and time required to reach the stationary state (bottom row), for different combinations of $\frac{\lambda}{\gamma}$ (from left to right 1, 2, 4, 10, respectively) and $\xi=0.5$. White spaces show that no stationary state is reached during the period of observation.}
 \label{fig:heatmap-w-P-x0.5}
\end{figure*} 

\subsection{Information Campaign (Social Immunization)}
 
We now show how a collaborative information campaign could help in contrasting the spread of the disease, following the model we presented in the last section. We initialize the scenario by distributing the immunizing information to 1\% of the population, randomly chosen regardless of their location. These people will be informed and will be instructed to spread the information. In other words, we assume that they will contact their social connections, according to the call matrix. 

In Fig.~\ref{fig:heatmap-wP-x} we show the density plots describing $i^\infty$ and $\tau$ for various values of $r_0$ (, for a subset of scenarios where $\omega=\psi$, i.e., when the information that spreads among the population has the same chance to immunize a person and to involve the person in the spreading process. This is consistent with a scenario where the same set of people who become aware also become immunized by the information they have received. Blank squares show that a stationary state was not reached for the corresponding set of parameters. The figure shows how contagious ($\omega$=$\psi$) the immunizing information  has to be with respect to how often people ``forget'' ($\xi$) in order to slow down the disease considerably and to reduce the endemic cases. When $\omega=\psi=0$ we fall back to the model without information spreading, and the value of $\xi$ does not affect $i^\infty$ and $\tau$. For $\omega=\psi>0$ and $\xi=0$ the fraction of infected population goes to zero in all cases, because the number of people aware of the information does not decrease, thus increasing the number of new immunized individuals at each step. We can notice that even for low values of participation $\omega$ and for information that gives temporary immunization ($\psi > 0$), the final fraction of infected individuals is considerably lower than in the case where no countermeasures are taken.

In Figs.~\ref{fig:heatmap-w-P-x0.0} and \ref{fig:heatmap-w-P-x0.5} we show the density plots for $\omega$ and $\psi$ when $\xi$ is constant. In particular, we analyze the scenario for $\xi=0$ (Fig.~\ref{fig:heatmap-w-P-x0.0}), which represents for example a scenario where the immunizing information is about vaccination campaigns (individuals who have been administered vaccination do not lose immunity). For every combination of parameters we have absence of endemic state even with the highest considered value of $r_0$. The two parameters that represent how individuals are likely to get involved both in the immunization and in the information spreading ($\omega$ and $\psi$) seem to have the same impact on the delay of the infection.

The value $\xi=0.5$ (Fig.~\ref{fig:heatmap-w-P-x0.5}) describes the scenario when the information is about a good practice (e.g., boiling water, using mosquito nets, etc.), which loses its effectiveness or it is stopped being used by a person with rate $\xi$. For this case we can notice that the fraction of infected population is independent from $\psi$, as rows in the density plot are of the same color. This suggests that, for this scenario, the rate at which people lose immunity does not affect the size of the endemic state.

%

%% file: conclusion.tex
\section{Conclusions}\label{sec:conclusion}

In this paper we have presented a model that describes the spreading of disease in a population where individuals move between geographic areas, extracted from cellular network records. We have showed the evolution of the disease and we have evaluated two types of countermeasures, namely the quarantine of central geographic areas and a collaborative ``viral'' information campaign among the population, by inferring the underlying social structure from the call records.


Our future research agenda includes the investigation of analytical aspects of the model, such as the derivation of the critical reproductive ratio $R_0$, i.e., the value that corresponds to the transition between an endemic and an endemic-free infection. Currently, the model is based on the assumption of a static mobility matrix: our goal is to refine the model by introducing time-dependent matrices, also exploring the application of the recent theoretical results related to temporal networks. We also plan to refine the model introducing specific contact rates for each metapopulation, potentially based on more fine-grained information about the number of encounters and the number of calls of each individual, if available. Finally, we plan to explore hybrid countermeasures, such as concurrent partial restrictions of mobility and targeted information campaigns.